\documentclass[twocolumn,showpacs,preprintnumbers,superscriptaddress,amsmath,amssymb,aps,pre]{revtex4-1}

\usepackage{graphicx}
\usepackage{dcolumn}
\usepackage{bm}
\usepackage{color}
\usepackage{longtable}
\usepackage{CJK}

\begin{document}
\begin{CJK*}{GBK}{Song} 


\title{Nonuniversal distributions of stock returns in an emerging market}

\author{Guo-Hua Mu}
 \affiliation{School of Business, East China University of Science and Technology, Shanghai 200237, China} %
 \affiliation{School of Science, East China University of Science and Technology, Shanghai 200237, China} %
 \affiliation{Research Center for Econophysics, East China University of Science and Technology, Shanghai 200237, China} %
\author{Wei-Xing Zhou}
 \email{wxzhou@ecust.edu.cn}
 \affiliation{School of Business, East China University of Science and Technology, Shanghai 200237, China} %
 \affiliation{School of Science, East China University of Science and Technology, Shanghai 200237, China} %
 \affiliation{Research Center for Econophysics, East China University of Science and Technology, Shanghai 200237, China} %
 \affiliation{Engineering Research Center of Process Systems Engineering (Ministry of Education), East China University of Science and Technology, Shanghai 200237, China}
 \affiliation{Research Center on Fictitious Economics \& Data Science, Chinese Academy of Sciences, Beijing 100080, China}

\date{\today}

\begin{abstract}
There is convincing evidence showing that the probability distributions of stock returns in mature markets exhibit power-law tails and both the positive and negative tails conform to the inverse cubic law. It supports the possibility that the tail exponents are universal at least for mature markets in the sense that they do not depend on stock market, industry sector, and market capitalization. We investigate the distributions of one-minute intraday returns of all the A-share stocks traded in the Chinese stock market, which is the largest emerging market in the world. We find that the returns can be well fitted by the $q$-Gaussian distribution and the tails have power-law relaxations with the exponents fluctuating around $\alpha=3$ and being well outside the L\'evy stable regime for individual stocks. We provide statistically significant evidence showing that the exponents logarithmically decrease with the turnover rate and increase with the market capitalization, and find that the market capitalization has a greater impact on the tail exponent than the turnover rate. Our findings indicate that the intraday return distributions are not universal in emerging stock markets.
\end{abstract}


\maketitle

\end{CJK*}

\section{Introduction}
\label{s1:Introduction}

The logarithmic return of stock price $S(t)$ over a time interval $\Delta{t}$ is defined as follows,
\begin{equation}
 r(t)=\ln S(t)- \ln S(t-\Delta{t}).
 \label{Eq:rt}
\end{equation}
The form of the distribution of asset price fluctuations plays crucial roles in asset pricing and risk management \cite{Mantegna-Stanley-2000,Bouchaud-Potters-2000,Malevergne-Sornette-2006}. Early empirical and theoretical works, which can be traced back to Bachelier in 1900 \cite{Bachelier-1900}, argue that asset prices follow geometric Brownian motions, i.e., the returns are normally distributed \cite{Bachelier-1900,Osborne-1959a-OR,*Black-Scholes-1973-JPE}. More than half a century after Bachelier's work, Mandelbrot finds that incomes and
speculative price returns follow the Pareto-L{\'{e}}vy distribution, which has power-law tails
\begin{equation}
 f(|r|) \sim |r|^{-\alpha-1},
 \label{Eq:PL}
\end{equation}
whose exponents $1<\alpha<2$ \cite{Mandelbrot-1960-IER,*Mandelbrot-1961-Em,*Mandelbrot-1963-IER,*Mandelbrot-1963-JPE,*Mandelbrot-1963-JB}. The picture of Paretian markets soon becomes the mainstream \cite{Fama-1965-JB,*Mantegna-Stanley-1995-Nature}.

In recent years, the Boston school shows that the tail distributions of returns of many stock indexes and stock prices for the USA markets exhibit an inverse cubic law, where the power-law exponents are close to $\alpha=3$ \cite{Gopikrishnan-Meyer-Amaral-Stanley-1998-EPJB,Gopikrishnan-Plerou-Amaral-Meyer-Stanley-1999-PRE,Plerou-Gopikrishnan-Amaral-Meyer-Stanley-1999-PRE}. A lot of empirical investigations have been conducted on financial returns at different time scales $\Delta{t}$ in different stock markets over different time periods, in which the distributions vary from exponential to stretched exponential to power-law fat-tailed
\cite{Laherrere-Sornette-1998-EPJB,Makowiec-Gnacinski-2001-APP,*Bertram-2004-PA,*Matia-Pal-Salunkay-Stanley-2004-EPL,*Yan-Zhang-Zhang-Tang-2005-PA,*Coronel-Hernandez-2005-PA,Qiu-Zheng-Ren-Trimper-2007-PA,*Drozdz-Forczek-Kwapien-Oswicimka-Rak-2007-PA,*Pan-Sinha-2007-EPL,*Pan-Sinha-2008-PA,*Tabak-Takami-Cajueiro-Petitiniga-2009-PA,*Eryigit-Cukur-Eryigit-2009-PA,*Jiang-Li-Cai-Wang-2009-PA,Queiros-2005-QF,*Queiros-Moyano-deSouza-Tsallis-2007-EPJB,Zhang-Zhang-Kleinert-2007-PA,Gu-Chen-Zhou-2008a-PA,Fuentes-Gerig-Vicente-2009-PLoS1,*Gerig-Vicente-Fuentes-2009-PRE}. This miscellaneous phenomenon is rational since the tails are expected to evolve from power law at small time scale to Gaussian at large scale
\cite{Ghashghaie-Breymann-Peinke-Talkner-Dodge-1996-Nature}, according to the variational theory for turbulent signals
\cite{Castaing-Gagne-Hopfinger-1990-PD,*Castaing-Gagne-Marchand-1993-PD,*Castaing-Chabaud-Hebral-Naert-Peinke-1994-PB,*Castaing-1994-PD}, which has been confirmed by numerous studies \cite{Gopikrishnan-Plerou-Amaral-Meyer-Stanley-1999-PRE,Plerou-Gopikrishnan-Amaral-Meyer-Stanley-1999-PRE,Silva-Prange-Yakovenko-2004-PA,*Kiyono-Struzik-Yamamoto-2006-PRL,*Wang-Hui-2001-JEPB,*Lee-Lee-2004-JKPS}.
Note that the stretched exponential distribution serves as a bridge between exponential and power-law distributions \cite{Laherrere-Sornette-1998-EPJB,*Sornette-2004}. Under certain conditions the stretched exponential density approaches to a power law, which is also supported by empirical evidence \cite{Malevergne-Pisarenko-Sornette-2005-QF,*Malevergne-Pisarenko-Sornette-2006-AFE,Malevergne-Sornette-2006}.

For intraday returns, the prevailing point is that the distributions have power-law tails outside the L{\'{e}}vy stable regime. Many efforts have been made to explain this power-law behavior. Base on ultra-high-frequency data, some researchers argue that large price fluctuations are caused by large trading volume \cite{Gabaix-Gopikrishnan-Plerou-Stanley-2003-Nature,*Gabaix-Gopikrishnan-Plerou-Stanley-2003-PA,*Gabaix-Gopikrishnan-Plerou-Stanley-2006-QJE,*Gabaix-Gopikrishnan-Plerou-Stanley-2007-JEEA,*Zhou-2007-XXX,*Gabaix-Gopikrishnan-Plerou-Stanley-2008-JEDC}, while some others submit that news and volume plays a minor role and the shortage of liquidity is the main cause \cite{Farmer-Gillemot-Lillo-Mike-Sen-2004-QF,*Weber-Rosenow-2006-QF,*Joulin-Lefevre-Grunberg-Bouchaud-2008-Wilmott}. The non-Gaussian behavior of stock prices can also be explained by the mixture-of-distribution hypothesis \cite{Clark-1973-Em,Fuentes-Gerig-Vicente-2009-PLoS1,*Gerig-Vicente-Fuentes-2009-PRE} or nonextensive statistical mechanics \cite{Queiros-2005-QF,*Queiros-Moyano-deSouza-Tsallis-2007-EPJB,Rak-Drozdz-Kwapien-2007-PA}. In addition, different classes of microscopic stock market models are also able to reproduce the power-law tailed distributions \cite{Challet-Marsili-Zhang-2005,*Preis-Golke-Paul-Schneider-2006-EPL,*Preis-Golke-Paul-Schneider-2007-PRE,*Mike-Farmer-2008-JEDC,*Gu-Zhou-2009-EPJB,*Gu-Zhou-2009-EPL}.

Despite the variation of tail exponents reported in the econophysics literature, there is convincing evidence showing that the tail exponents for stocks in western mature markets are possibly universal based on a careful study of the trade-by-trade data of 1000 major USA stocks, 85 major stocks traded on the London Stock Exchange which form part of the FTSE 100 index, and 13 major stocks traded on the Paris Bourse that form part of the CAC 40 index \cite{Plerou-Stanley-2008-PRE,*Stanley-Plerou-Gabaix-2008-PA}. It shows that the exponent values are $\alpha\approx3$ for the three markets and do not display variations with respect to market capitalization or industry sector.

In this paper, we investigate the distributions of one-minute intraday returns of all the A-share stocks traded in the Chinese stock market, which is the largest emerging market in the world. Our aim is to test the possible dependence of the tail exponents on the turnover rate and market capitalization. There is statistically significant evidence showing that the exponents increase with market capitalization and decrease with turnover rate. It indicates that, different from developed markets, the intraday return distributions are not universal in emerging stock markets.

\section{Data sets}
\label{S1:Data}

We employ a nice tick-by-tick database of the A-share stocks for all companies traded on the Shenzhen Stock Exchange and the Shanghai Stock Exchange from January 2004 to June 2006. The data were recorded based on the market quotes disposed to all traders in every six to eight seconds, which are different from the ultrahigh-frequency data reconstructed from order flows \cite{Gu-Chen-Zhou-2007-EPJB}.

We compute the one-minute intraday returns for each stock. We emphasize that the {\em{intraday}} returns are calculated within individual trading days to eliminate the overnight effect \cite{Zhang-Zhang-Kleinert-2007-PA,Wang-Shieh-Havlin-Stanley-2009-PRE}. For each stock, the one-minute returns are normalized so that the mean is 0 and the variance is 1.

\section{Dependence of return distributions on turnover rate}
\label{S1:Value}

We investigate the possible impact of the turnover rate on the distribution of stock returns, especially on the power-law tail exponent. For each stock, we first calculate the total traded value of all trades in each minute
\begin{equation}
 v = \sum_{i=1}^n v_i p_i,
 \label{Eq:V}
\end{equation}
where $n$ is the number of trades, $v_i$ is the size of the $i$-th trade, and $p_i$ is the price of the $i$-th trade. The 1-min average traded value $\langle{v}\rangle$ is the average over all 1-min time interval for the stock and the 1-min turnover rate is calculated by the ratio of average traded values to market capitalization.

\begin{figure}[htb]
\centering
\includegraphics[width=7cm]{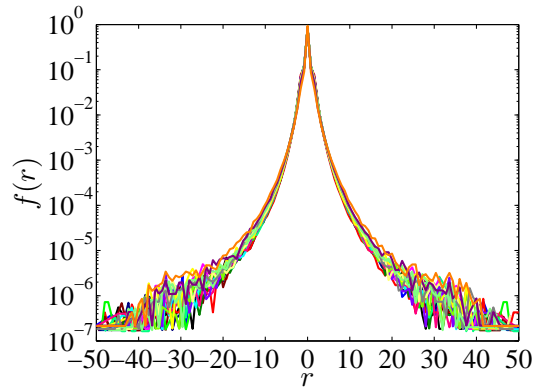}
\caption{\label{Fig:Turnove:PDF} (Color online) Empirical distributions of 1-min stock returns for the 20 groups partitioned based on the average turnover rate.}
\end{figure}

We sort all the stocks according to their 1-min turnover rates and partition them into 20 groups, ensuring that the groups contain almost identical number of stocks. The 1-min returns of the stocks in the same group are pooled as one sample. Figure \ref{Fig:Turnove:PDF} shows the empirical distributions of stock returns for the 20 groups. We find that all these distributions share a qualitatively similar shape with fat tails. A careful scrutiny shows that the curves are not smooth around $r=0$. This is due to the fact that the prices of individual stocks have a tick size \cite{Plerou-Gopikrishnan-Amaral-Meyer-Stanley-1999-PRE}. This phenomenon disappears when we investigate stock indices \cite{Gopikrishnan-Plerou-Amaral-Meyer-Stanley-1999-PRE}.

\begin{figure}[htb]
\centering
\includegraphics[width=7cm]{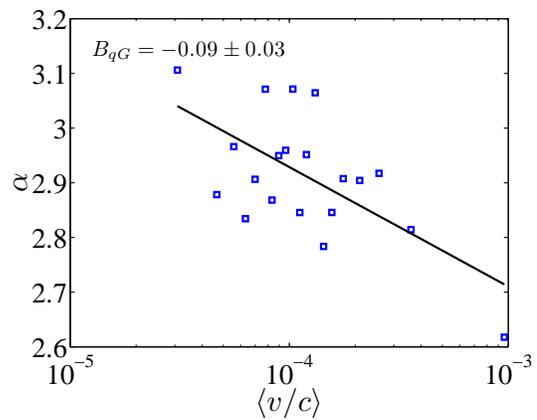}
\caption{\label{Fig:Turnove:qG} (Color online) Dependence of the exponent $\alpha$ estimated from the $q$-Gaussian function with respect to the average turnover rate $\langle{v/c}\rangle$. The solid line gives the logarithmic regression $\alpha=A+B\ln\langle{v/c}\rangle$, where $B=-0.09\pm0.03$ and $R^2= 0.40$.}
\end{figure}

For the 1-min returns of Chinese stocks, the distributions can be well fitted by Student's t-distribution \cite{Gu-Chen-Zhou-2008a-PA}
\begin{equation}
 f_{t}(r)= \sqrt{L} n^{n/2}/B\left(\frac{1}{2},\frac{n}{2}\right) \left(\alpha+Lr^2\right)^{-(\alpha+1)/2},
 \label{Eq:qG}
\end{equation}
where $B(\cdot, \cdot)$ is the ``beta function'', $L$ is the scale parameter, and $\alpha$ is the degrees of freedom. The Student distribution is also known as the $q$-Gaussian distribution in nonextensive statistical mechanics \cite{Tsallis-1988-JSP,Queiros-2005-QF,*Queiros-Moyano-deSouza-Tsallis-2007-EPJB}. We fit the 20 curves using Eq.~(\ref{Eq:qG}) and plot the estimated exponent $\alpha$ as a function of $\langle{v/c}\rangle$ in Fig.~\ref{Fig:Turnove:qG}. We adopt a logarithmic form
\begin{equation}
 \alpha=A+B\ln\langle{v/c}\rangle,
 \label{Eq:alpha:lnvc}
\end{equation}
and a linear regression gives $B=-0.09\pm0.03$, where the error is determined according to the standard t-test at the 5\% significance level. There is a decreasing trend between $\alpha$ and $\langle{v/c}\rangle$.

The estimates of the tail exponents based on the fitting of the $q$-Gaussian function might be biased since the bulk of the distribution with not large returns has dominating impact on the objective function of the fitting. We thus use a well designed method proposed for estimating tail exponents, which is known as the CSN method \cite{Clauset-Shalizi-Newman-2009-SIAMR}. We briefly review the idea of the CSN method for positive returns, which can be easily extended to negative returns. The CNS method has a promising advantage to determine the cutoff $r_{\min}$ in an objective way. After the cutoff $r_{\min}$ is determined, the tail exponent of the returns $r\geqslant r_{\min}$ can be determined based on the maximal likelihood estimation
\cite{Clauset-Shalizi-Newman-2009-SIAMR},
\begin{equation}
 \alpha = N \sum_{j=1}^{N} \ln \frac{r_j}{r_{\min}-0.5},
 \label{Eq:CSN}
\end{equation}
where $N$ is the number of returns $r$ that are no less than $r_{\min}$. In order to determine $r_{\min}$, one needs to scan different values of $r$ to determine the corresponding parameters and obtain the Kolmogorov-Smirnov or KS statistics. The optimal cutoff $r_{\min}$ is the one that minimizes the KS statistic.

\begin{figure}[htb]
\centering
\includegraphics[width=7cm]{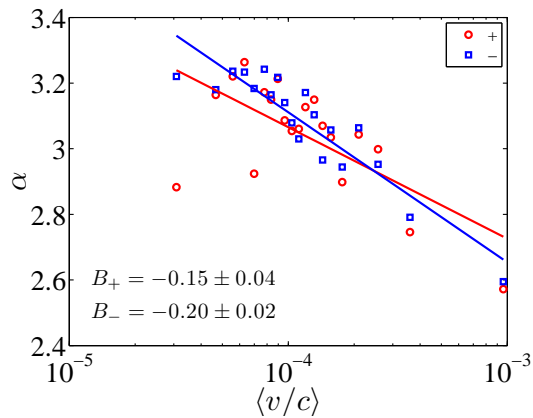}
\caption{\label{Fig:Turnove:alpha:20bins} (Color online) Dependence of the tail exponents of positive and negative returns for the 20 groups as a function of the 1-min average turnover rate $\langle{v/c}\rangle$. The solid lines are the best fits of $\alpha=A+B\ln{\langle{v/c}\rangle}$, where $B=-0.15\pm0.04$ for the positive tails with the $p$-value equaling to 0 and $R^2= 0.45$, and $B=-0.20\pm0.02$ for the negative tails with the $p$-value equaling to 0 and $R^2= 0.84$, respectively.}
\end{figure}

Figure \ref{Fig:Turnove:alpha:20bins} shows the dependence of the tail exponents of positive and negative returns for the 20 groups as a function of the 1-min average trading value. We fit the data to the logarithmic function (\ref{Eq:alpha:lnvc}) and obtain that $B=-0.15\pm0.04$ for the positive tails with $R^2= 0.45$ and $B=-0.20\pm0.02$ for the negative tails with $R^2= 0.84$, respectively. We find that the tail exponents decrease with the turnover rate. For the positive return curve, there are two outliers that deviate remarkably from the linear trend. If we discard these two points, the results for positive and negative returns are very similar.

\begin{figure}[htb]
\centering
\includegraphics[width=7cm]{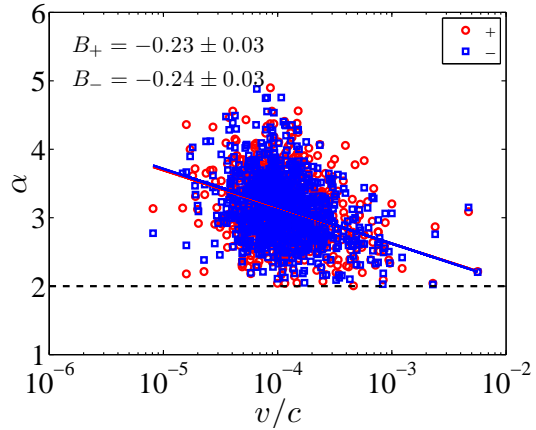}
\caption{\label{Fig:Turnove:alpha:EveryStock} (Color online) Dependence of the tail exponents of returns for individual stocks as a function of the 1-min average turnover rate $\langle{v/c}\rangle$. The solid lines are the best fits of $\alpha=A+B\ln{\langle{v/c}\rangle}$, where $B=-0.23\pm0.03$ for the positive tails with the $p$-value equaling to 0 and $R^2= 0.07$ and $B=-0.24\pm0.03$ for the negative tails with the $p$-value equaling to 0 and $R^2= 0.08$, respectively. The dashed line is $\alpha=2$.}
\end{figure}

We also apply the CSN method to the returns of individual stocks and the exponents of positive and negative tails are calculated. The dependence of the tail exponents of returns for individual stocks as a function of the 1-min turnover rate is presented in Fig.~\ref{Fig:Turnove:alpha:EveryStock}. Fitting the data to Eq.~(\ref{Eq:alpha:lnvc}), we have $B=-0.23\pm0.03$ for the positive tails with $R^2= 0.07$ and $B=-0.24\pm0.03$ for the negative tails with $R^2= 0.08$, respectively. Although the R-square values are small, the slopes $B$ are significantly different from 0, which implies that the tail exponents do depend on the turnover rate. In addition, we find that the positive and negative tails are roughly symmetric.

All the results presented above show that the tails are fatter with smaller exponents $\alpha$ for larger turnover rates. This is consistent with the conventional wisdom that a stock with high turnover rate is riskier and has higher volatility. Usually, stocks with small market capitalization have higher turnover rates, which is especially true for the Chinese stock market. Therefore, it is possible that market capitalization may also have an important impact on the heaviness of the return tails.

\section{Dependence of return distribution on market capitalization}
\label{S1:Cap}

We now investigate the possible impact of market capitalization on the distribution of stock returns, especially on the power-law tail exponent. We note that the ownership of stock shares is split and only about one-third of the total outstanding shares are tradable in the market during the time period under investigation. The market capitalization is calculated as the product of the outstanding tradable shares and the price at the beginning of the database records.

\begin{figure}[htb]
\centering
\includegraphics[width=7cm]{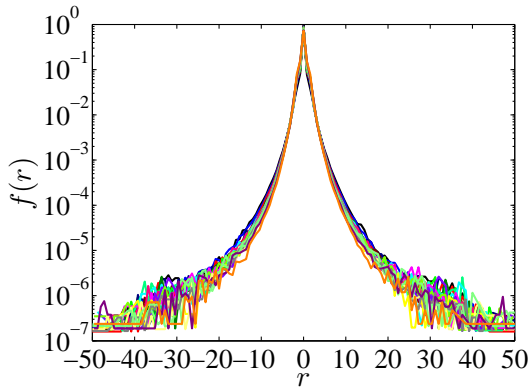}
\caption{\label{Fig:Cap:PDF} (Color online) Empirical distributions of 1-min stock returns for the 20 groups partitioned based on the average market capitalization.}
\end{figure}

We sort all the stocks according to their market capitalization and partition them into 20 groups, ensuring that the groups contain almost identical number of stocks. The 1-min returns of the stocks in the same group are pooled as one sample. Figure \ref{Fig:Cap:PDF} shows the empirical distributions of stock returns for the 20 groups with different average market capitalization. It is found that the shape of the curves is very similar to that in Fig.~\ref{Fig:Turnove:PDF}. Comparing Fig.~\ref{Fig:Cap:PDF} and Fig.~\ref{Fig:Turnove:PDF}, we observe that the curves in Fig.~\ref{Fig:Cap:PDF} are less collapsed with each other. It can be conjectured that the correlation between tail exponent and market capitalization is stronger than the turnover rate, which is exactly the case as we will show below.

\begin{figure}[htb]
\centering
\includegraphics[width=7cm]{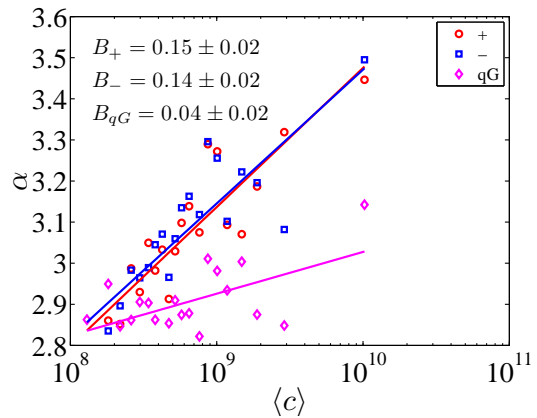}
\caption{\label{Fig:Cap:alpha:20bins} (Color online) Dependence of the tail exponents of returns for the 20 groups as a function of the average market capitalization. The tail exponents are estimated from the $q$-Gaussian model and the CSN method. The three sets of data points are fitted to the logarithmic function (\ref{Eq:alpha:lnC}). We obtain that $B=0.04\pm0.02$ with $R^2=0.32$ for the $q$-Gaussian model, $B=0.15\pm0.02$ with $R^2=0.80$ for the positive tails, and $B=0.14\pm0.02$ with $R^2=0.76$ for the negative tails, respectively.}
\end{figure}

We estimate the tail exponents using the $q$-Gaussian model and the CSN method as in Sec.~\ref{S1:Value}. For each group of stocks, we obtain three exponents. Figure \ref{Fig:Cap:alpha:20bins} illustrates the dependence of the tail exponents of 1-min returns for the 20 groups as a function of the average market capitalization. We observe that all the three exponents exhibit an increasing linear trend in semi-logarithmic coordinates. To fit the three curves, we adopt a logarithmic function
\begin{equation}
 \alpha=A+B\ln\langle{c}\rangle.
 \label{Eq:alpha:lnC}
\end{equation}
Linear regressions of $\alpha$ with respect to $\ln{c}$ give that $B=0.04\pm0.02$ with $R^2=0.32$ for the $q$-Gaussian model, $B=0.15\pm0.02$ with $R^2=0.80$ for the positive tails, and $B=0.14\pm0.02$ with $R^2=0.76$ for the negative tails, respectively. Comparing with the results in Sec.~\ref{S1:Cap}, the R-square values in the present case are much larger and the slopes are all significantly different from 0 according to the t-tests. In other words, the tails for small-cap stocks are fatter than large-cap stocks. This finding is consistent with the conventional wisdom in finance that small-cap stocks are riskier and have more occurrences of large price fluctuations.

\begin{figure}[htb]
\centering
\includegraphics[width=7cm]{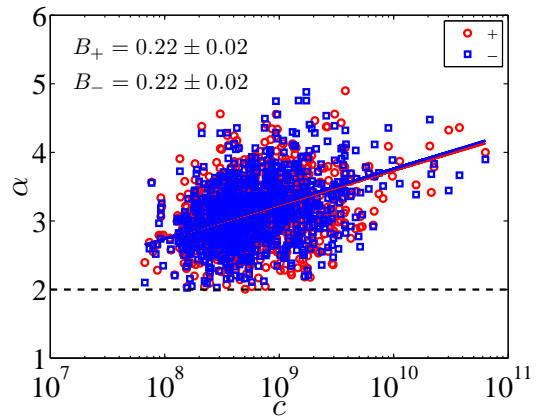}
\caption{\label{Fig:Cap:alpha:EveryStock} (Color online) Dependence of the tail exponents of returns for individual stocks as a function of the market capitalization. The solid lines are the best fits of $\alpha=A+B\ln{c}$, where $B=0.22\pm0.02$ for the positive tails  with $R^2= 0.10$ and $B=0.22\pm0.03$ for the negative tails with $R^2= 0.12$, respectively.}
\end{figure}

We also apply the CSN method to the returns of individual stocks and the exponents of positive and negative tails are calculated. Figure~\ref{Fig:Cap:alpha:EveryStock} shows the dependence of the tail exponents of returns for individual stocks as a function of the average market capitalization. Fitting the data to Eq.~(\ref{Eq:alpha:lnC}) where $\langle{c}\rangle$ is replaced by $c$, we have $B=0.22\pm0.02$ for the positive tails  with $R^2= 0.10$ and $B=0.22\pm0.03$ for the negative tails with $R^2= 0.12$, respectively. We find that the slopes $B$ are significantly different from 0. We also observe that the $B$ values in Fig.~\ref{Fig:Cap:alpha:EveryStock} for individual stocks are larger than those in Fig.~\ref{Fig:Cap:alpha:20bins} for grouped stocks. In addition, we find that the positive and negative tails are roughly symmetric.

\section{Bivariate regression}
\label{S1:BiVarRegress}

In Sec.~\ref{S1:Value} and Sec.~\ref{S1:Cap}, we have shown that the tail exponents are dependent of the average turnover rates and the market capitalization in logarithmic forms. It is thus natural to combine these results and suggest a bivariate logarithmic function as follows
\begin{equation}
 \alpha = A + B_{v/c}\ln\langle{v/c}\rangle + B_{c}\ln c.
 \label{Eq:alpha:lnv:lnC}
\end{equation}
This test can only be done for individual stocks since the two grouping methods in the previous sections are different.

We perform linear least-squares regressions for positive and negative tails. The estimated parameters for positive tails are obtained that $A = -1.13\pm0.41$, $B_{v/c} =-0.07\pm0.04$, and $B_c = 0.18\pm0.03$, whose R-square is $R^2 = 0.11$ and the $p$-values are 0.006, 0.077, and 0, respectively. For negative tails, the parameters are $A =-1.23\pm0.39$, $B_{v/c} = -0.06\pm0.03$, and $B_c=0.19\pm0.03$, whose R-square is $R^2 = 0.13$ and the $p$-values are 0.001, 0.066, and 0, respectively. A comparison of the three regression models expressed in Eqs. (\ref{Eq:alpha:lnvc}), (\ref{Eq:alpha:lnC}) and (\ref{Eq:alpha:lnv:lnC}) is presented in Table \ref{TB:Models}.

\begin{table}[htb]
    \centering
    \caption{Comparison of the three regression models presented in Eqs. (\ref{Eq:alpha:lnvc}), (\ref{Eq:alpha:lnC}) and (\ref{Eq:alpha:lnv:lnC}). The number of stars in the superscript of a coefficient indicates its significance level: $***$ for 1\% and $*$ for 10\%.}
    \medskip
    \label{TB:Models}
    \begin{tabular}{ccccccccccccccccccccccccc}
    \hline\hline
    &\multicolumn{4}{@{\extracolsep\fill}c}{Positive tail}&&\multicolumn{4}{@{\extracolsep\fill}c}{Negative tail}\\
    \cline{3-6} \cline{8-11}
            Eqn              && $B_{v/c}      $ & $B_{c}      $ && $R^2$ && $B_{v/c}       $ & $B_{c}        $ && $R^2$ \\\hline
    (\ref{Eq:alpha:lnvc})     && $-0.23^{***}  $ & $  /~~~     $ && 0.07  && $-0.24^{***}   $ & $   /~~~      $ && 0.08\\%
    (\ref{Eq:alpha:lnC})     && $  /~~~       $ & $0.22^{***} $ && 0.10  && $    /~~~      $ & $0.22^{***}   $ && 0.12\\%
    (\ref{Eq:alpha:lnv:lnC}) && $-0.07^{*~~~} $ & $0.18^{***} $ && 0.11  && $-0.06^{*~~~}  $ & $0.19^{***}   $ && 0.13\\%
   \hline\hline
    \end{tabular}
\end{table}

The bivariate regression results are consistent with those in the univariate regressions. That is, the tail exponents increase with market capitalization and decrease with turnover rate. For both univariate model (\ref{Eq:alpha:lnC}) and bivariate model (\ref{Eq:alpha:lnv:lnC}), the coefficients of market capitalization are significant at the 1\% level.  The coefficients of turnover rate are significant at the 1\% level in the univariate model (\ref{Eq:alpha:lnvc}) and significant only at the 10\% level in the bivariate model (\ref{Eq:alpha:lnv:lnC}). In addition, the introducing of $\ln{c}$ in model (\ref{Eq:alpha:lnvc}) improves its explanatory power (characterized by $R^2$) by 4\% to 5\%, while introducing $\ln\langle{v/c}\rangle$ in model (\ref{Eq:alpha:lnC}) improves its explanatory power by 1\%. All these findings imply that market capitalization is a more significant influencing factor of the tail heaviness.

A simple manipulation of Eq.~(\ref{Eq:alpha:lnv:lnC}) reads
\begin{equation}
 \alpha = A + B_{v/c}\ln\langle{v}\rangle + (B_{c}-B_{v/c})\ln c,
 \label{Eq:alpha:lnv:lnC:bis}
\end{equation}
which means that the average traded value might also have an impact on the tail heaviness. We have investigated the univariate model by posing $B_{c}-B_{v/c}=0$ in Eq.~(\ref{Eq:alpha:lnv:lnC:bis}), that is
\begin{equation}
 \alpha = A + B\ln\langle{v}\rangle,
 \label{Eq:alpha:lnv}
\end{equation}
We partition the stocks into 20 groups according to their 1-min average traded values. From the $q$-Gaussian model, we find that $B=-0.06\pm0.04$ with $R^2= 0.12$. In contrast, the CSN method gives that $B=0.01\pm0.02$ with $R^2=0.02$ for positive tails and $B=0.02\pm0.02$ with $R^2=0.03$ for negative tails. For individual stocks, we have $B=0.16\pm0.03$ with $R^2=0.03$ for positive tails and $B=0.17\pm0.03$ with $R^2=0.04$ for negative tails. Different from the results of turnover rate and market capitalization, the results of average traded value from different method are inconsistent with each other.

\section{Conclusion}
\label{S1:conclusion}

We have investigated the distributions of one-minute intraday returns of all the A-share stocks traded in the Chinese stock market, which is the largest emerging market in the world. The returns are standardized to have zero mean and unit variance. We studied the possible impact of the turnover rate and the market capitalization on the return distributions with special attention paid to the tail behavior.

For individual stocks, the returns are found to have power-law tails. Basically, the tail exponents fluctuate around $\alpha=3$, ranging from $\alpha=2$ to $\alpha=5$. It indicates that the return distributions of individual Chinese stocks are well outside the L\'evy stable regime. When the stocks are grouped according to their turnover rates, market capitalization or traded values, the returns in each group can be well fitted by the $q$-Gaussian formula.

We found from different methods that the tail exponents logarithmically decrease with the turnover rate and increase with the market capitalization and the market capitalization has a greater impact than the turnover rate. These observations are consistent with the fact that a stock is riskier and has fatter tails when its capitalization is small and it has higher turnover rate. However, we did not find convincing evidence for the impact of traded value on the tail behavior. We conclude that the intraday return distributions are not universal in emerging stock markets, which is different from the universal tail distribution for stocks in developed western stock markets.

\begin{acknowledgements}

We acknowledge financial support from the ``Shu Guang'' project (2008SG29) and the ``Chen Guang'' project (2008CG37) sponsored by Shanghai Municipal Education Commission and Shanghai Education Development Foundation and the Program for New Century Excellent Talents in University (NCET-07-0288).
\end{acknowledgements}

\bibliography{E:/Papers/Auxiliary/Bibliography}

\end{document}